\documentclass[11pt]{article}

\usepackage[utf8]{inputenc}
\usepackage[T1]{fontenc}
\usepackage[margin=1in]{geometry}
\usepackage{booktabs}
\usepackage{array}
\usepackage{amsmath}
\usepackage{graphicx}
\usepackage{xcolor}
\usepackage{float}          
\usepackage{enumitem}       
\usepackage{tcolorbox}  
\usepackage{tikz}
\usepackage{url}
\usepackage[colorlinks=true,allcolors=blue]{hyperref}
\usepackage{graphicx}
\usepackage{xcolor}
\usepackage{tikz}
\usepackage{tcolorbox}
\usepackage{subcaption} 
\usepackage{float} 

\usepackage{booktabs}       
\usepackage{tabularx}       
\usepackage{tabularray}     
\usepackage{multirow}       
\usepackage{colortbl}       
\usepackage{threeparttable} 
\usepackage{array}          
\usepackage{makecell}       

\usepackage{pifont}         
\usepackage{wasysym}        
\usepackage{amssymb}       

\usepackage{enumitem}       

\definecolor{Gray}{gray}{0.92}

\newcommand{\cmark}{\textcolor{black}{\ding{51}}} 

\newcommand{\fullcirc}{\CIRCLE}
\newcommand{\halfcirc}{\LEFTcircle}
\newcommand{\emptycirc}{\Circle}

\title{Extracting and Verifying Illicit Bitcoin Addresses from Underground Forum Discussions}

\author{
{\rm Abdoul Nasser Hassane Amadou$^{\ast}$}\\
\phantom{x}
\and
{\rm Arnaud Legout$^{\dagger}$}\\
\phantom{x}
\and
{\rm Imane Fouad$^{\ast}$}\\
\phantom{x}
\and
{\rm Konstantin Avrachenkov$^{\dagger}$ \quad Anas Motii$^{\ast}$}\\
\phantom{x}
\\[6pt]
\footnotesize
$^{\ast}$College of Computing, Mohammed VI Polytechnic University (UM6P), Morocco \\ \quad
$^{\dagger}$Inria, Universit\'e C\^ote d'Azur, France
}

\begin{document}
\maketitle

\begin{abstract}
Existing labeled Bitcoin datasets are largely derived from community-reported abuse, blockchain heuristics, incident-specific collections, or proprietary labeling processes. Their construction methods are rarely publicly reproducible and often provide limited evidence that an address was directly involved in illicit activity.  We present a reproducible pipeline for constructing evidence-backed Bitcoin labels from HackForums, an underground cybercrime forum with fifteen years of archived activity. The pipeline combines LLM-assisted screening, expert review, and on-chain validation to identify Bitcoin addresses explicitly associated with illicit transactions discussed on the forum. Each released label is supported by contextual evidence from underground discussions and validated on-chain. The resulting dataset contains 2,438 manually verified illicit Bitcoin addresses spanning 2010-2024 and twelve cybercrime categories assigned during LLM screening. We release the dataset, temporal metadata, and the complete extraction pipeline to support reproducible research on cryptocurrency-facilitated cybercrime.
\end{abstract}

\section{Introduction}
\label{sec:intro}

Cryptocurrencies have become a primary payment mechanism for cybercrime, including ransomware, malware-as-a-service, underground marketplaces, fraud, and money laundering. Consequently, identifying cryptocurrency addresses linked to illicit activity is essential for security researchers, law enforcement agencies, and financial institutions.

Despite extensive research, establishing the provenance of labels in existing Bitcoin datasets remains a fundamental challenge. Many datasets are constructed from community reports, blockchain heuristics, incident-specific collections, or proprietary intelligence sources. These data sources provide varying levels of evidential support, and their labeling processes are often difficult or impossible to reproduce. As a result, many public datasets lack verifiable evidence directly linking addresses to illicit activity, making it difficult to assess label quality and fairly compare detection methods.

To address this limitation, we construct a dataset of manually verified illicit Bitcoin addresses extracted from HackForums, one of the largest and longest-running underground cybercrime forums. We extract and label cryptocurrency addresses from more than 42 million forum posts, identifying those explicitly linked to illicit activity. The pipeline first extracts candidate strings and verifies their on-chain transaction history. It then uses the large language model Qwen3-30B to screen for posts likely to describe illicit activity, substantially reducing the manual review effort. Every flagged address is subsequently examined by a domain expert, who reviews the posts containing the address and their surrounding context before assigning the final illicit or licit label.

The resulting dataset contains 2,438 manually verified illicit Bitcoin addresses spanning fifteen years of underground activity. Each address is supported by both contextual evidence from HackForums and on-chain validation, and is associated with one of twelve cybercrime categories assigned during the screening process, including fraud, malware, money laundering, hacking-for-hire, and fraudulent investment schemes. To support reproducible research, we publicly release the dataset together with its temporal metadata and the complete extraction pipeline.

Our contributions are summarized as follows:

\begin{itemize}[leftmargin=*]
    \item We present a reproducible pipeline for constructing evidence-backed Bitcoin labels from underground forums by combining on-chain validation, LLM-assisted candidate screening, and manual expert verification.

    \item We construct and publicly release a dataset of 2,438 manually verified illicit Bitcoin addresses collected from fifteen years of HackForums activity, together with their temporal metadata, cybercrime category annotations, and the complete extraction pipeline to support reproducible research on cryptocurrency-facilitated cybercrime.\footnote{Dataset available at: \url{https://github.com/osteoner/Hackforums-dataset}}
\end{itemize}
\section{Related Work}
Labeled Bitcoin datasets differ substantially in their label provenance, supporting evidence, and verification methodology. Table~\ref{tab:datasets_landscape} summarizes the current landscape, comparing dataset scale, coverage, and verification methodology.

The Elliptic dataset~\cite{weber2019anti} and its extensions~\cite{elmougy2023demystifying,bellei2024shape} provide hundreds of thousands of labeled addresses but rely on proprietary intelligence, preventing independent verification of their labels. In contrast, publicly verifiable datasets are typically constructed from documented incidents, such as ransomware campaigns~\cite{CONTI2018162,paquet2019ransomware,oosthoek2023tale}, Ponzi schemes~\cite{Bartoletti2018Ponzi}, exchange scams~\cite{XIA2020101993}, and giveaway scams~\cite{li2023double}. Although these datasets provide strong supporting evidence, they are generally limited to a single crime category and a narrow time period.

Recent studies have expanded coverage by aggregating multiple public datasets into larger benchmarks~\cite{akcora2020bitcoinheist,xiang2023babd,schnoering2025bitcoin}. While this increases label diversity, the resulting datasets inherit the labeling criteria, confidence levels, and provenance limitations of their constituent sources. Many also report substantially larger dataset sizes by expanding seed addresses through clustering heuristics, such as the multi-input and change-address heuristics, which can propagate labeling errors beyond the originally verified addresses. 

We address this gap by constructing a large-scale, evidence-backed dataset
of manually verified illicit Bitcoin addresses from HackForums. Our pipeline
combines contextual evidence from forum discussions with on-chain validation,
LLM-assisted candidate screening, and expert verification, and we publicly
release both the dataset and the complete extraction pipeline.

\begin{table*}[t]
\centering
\caption{Labeled Bitcoin datasets. \textit{Verif.} (Label Verifiability) follows a three-value scale: \textbf{\fullcirc~Full} (derivable from fully public sources); \textbf{\halfcirc~Partial} (some public sources, but key steps are undisclosed); \textbf{\emptycirc~None} (relies entirely on proprietary intelligence). \textit{MI}: Multi-Input Heuristic used; \textit{CA}: Change Address Heuristic used. $^{\dagger}$For our dataset, \#Total denotes the pool of on-chain-validated candidate addresses extracted from the forum, of which 2,438 are expert-verified illicit.}
\label{tab:datasets_landscape}
\resizebox{\textwidth}{!}{%
\begin{tabular}{@{}lllrrllll@{}}
\toprule
\textbf{Dataset} & \textbf{Venue (Year)} & \textbf{Verif.} & \textbf{\#Total} & \textbf{\#Illicit} & \textbf{Crime Scope} & \textbf{MI} & \textbf{CA} & \textbf{Public} \\
\midrule
Elliptic (Elliptic1) \cite{weber2019anti}          & KDD'19           & \emptycirc & 203{,}769        & 4{,}545        & Mixed                     & No  & No  & \cmark \\
Elliptic++ \cite{elmougy2023demystifying}          & KDD'23           & \emptycirc & 822{,}942 addr.  & 14{,}266     & Mixed                     & Yes  & No  & \cmark \\
Elliptic2 \cite{bellei2024shape}                   & KDD MLF'24       & \emptycirc & 121{,}810 subgr. & 2{,}763        & Money laundering                   & Yes & Yes & \cmark \\
BitcoinHeist \cite{akcora2020bitcoinheist} & IJCAI'20 & \fullcirc  & $\sim$2.9M & 41{,}413 & Ransomware (28 fam.) & Yes & No  & \cmark \\
BABD-13 \cite{xiang2023babd} & IEEE TIFS'24 & \halfcirc  & 544{,}462 & 140,274 & Blackmail, darknet, gambling, Gov blacklist, ML, Ponzi & No  & No  & \cmark \\
BTC Temporal Graph \cite{schnoering2025bitcoin}    & Sci. Data'25     & \fullcirc  & 252M nodes       & $\sim$13.5K    & Entity typing (broad)              & Yes & No  & \cmark \\
Conti et al. \cite{CONTI2018162}              & Comput. Secur.'18  & \fullcirc  & 4{,}022          & 4{,}022        & Ransomware (20 fam.)               & Yes & Yes & \cmark \\
Bartoletti et al. \cite{Bartoletti2018Ponzi}        & CVCBT'18         & \fullcirc  & 7{,}611          & 1{,}211        & Ponzi (32 schemes)                 & Yes & No  & \cmark \\
Paquet-Clouston et al. \cite{paquet2019ransomware} & J. Cybersecur.'19& \fullcirc  & 7{,}118          & 7{,}118        & Ransomware (35 fam.)               & Yes & No  & \cmark \\
Sextortion \cite{paquet2019Sextortion} & AFT'19 & \halfcirc & 12{,}533 & 12{,}533 & Sextortion spam (35) & Yes & No & \cmark \\
Exchange Scams \cite{XIA2020101993}        & Comput. Secur.'20& \fullcirc  & 2{,}020          & 182            & Exchange scams (94)                & No  & Yes & \cmark \\
Oosthoek et al. \cite{oosthoek2023tale} & CACM'23 & \fullcirc & 7{,}321 & 7{,}321 & Ransomware (87 fam.) & Yes & Yes & \cmark \\
Gray et al. (Conti) \cite{gray2022money} & eCrime'22 & \fullcirc  & 741 & 741 & Conti ransomware & Yes & No & \cmark \\
Giveaway Scams \cite{li2023double} & NDSS'23 & \fullcirc  & 2{,}266 & 2{,}266 & Crypto giveaway scams & No  & No  & \cmark \\
Cable et al. (Ransomwhere) \cite{cable2024showing} & eCrime'24 & \halfcirc & 1{,}013 & 1{,}013 & Ransomware & Yes & Yes & \cmark \\
HackForums & \textbf{(Ours'26)} & \fullcirc & 23{,}882$^{\dagger}$ & 2{,}438 & Mixed & No & No & \cmark \\
\bottomrule
\end{tabular}%
}
\end{table*}

\section{Methodology}
\begin{figure}
    \centering
    \includegraphics[width=1\linewidth]{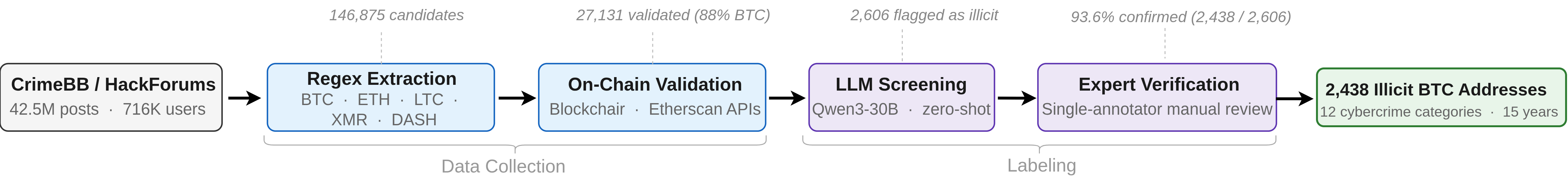}
    \caption{Extracting, validating, and expert-verifying illicit Bitcoin addresses from HackForums.}
    \label{fig:pipeline}
\end{figure}

Our pipeline consists of three stages: Bitcoin address extraction using regular expressions, on-chain validation, and a two-stage labeling process combining LLM-based screening with expert verification. Figure~\ref{fig:pipeline} illustrates the end-to-end architecture of this pipeline.

\subsection{Data Source}

We use the CrimeBB academic dataset~\cite{Pastrana2018CrimeBB}, a large archive of underground cybercrime forums. Specifically, we analyze the \textit{HackForums} snapshot dated 
2024-09-04, one of the largest and longest-running cybercrime forums. The snapshot contains more than 42 million posts across 212 boards (subforums), contributed by more than 700,000 users over a period of fifteen years. Owing to its scale and broad coverage of cybercrime discussions, HackForums provides a rich source of Bitcoin addresses associated with illicit activity. Table~\ref{tab:dataset-stats} summarizes the dataset.

\begin{table}[t]
\centering
\caption{Overview of the HackForums dataset statistics. \emph{Subforums} are the forum's topical boards (e.g., marketplace, hacking tutorials, and leaks). \emph{Users} and \emph{posts} denote the total number of registered accounts and forum messages, respectively. \emph{Threads} are topic-level discussions that group related posts. \emph{Contracts} are formal trade agreements between two forum members used to record marketplace transactions. \emph{Average post length} and \emph{average thread title length} denote the mean word counts of individual posts and thread titles, respectively.}
\label{tab:dataset-stats}
\vspace{5pt}
\begin{tabular}{@{}lr@{}}
\toprule
\textbf{Metric} & \textbf{Value} \\
\midrule
Number of subforums          & 212        \\
Number of users              & 716,058    \\
Number of posts              & 42,474,325 \\
Number of threads            & 4,148,196  \\
Number of contracts          & 348,340    \\
Average post length (words)  & 27.61      \\
Average thread title length (words)& 5.29       \\ 
\bottomrule
\end{tabular}
\end{table}

\subsection{Address Extraction}

We scan all forum posts and contract records for cryptocurrency addresses using regular expressions covering five cryptocurrencies commonly associated with cybercrime: Bitcoin (BTC), Ethereum (ETH), Litecoin (LTC), Dash (DASH), and Monero (XMR). The regex patterns are listed in Table~\ref{tab:address-regex}. To maximize recall, we retain all matches, including false positives such as cryptographic hashes, source code fragments, and transaction identifiers. This initial extraction yielded 146,875 candidate strings.

\begin{table}[h]
\centering
\caption{Regular expressions used to extract candidate cryptocurrency addresses.}
\label{tab:address-regex}
\vspace{5pt}
\begin{tabularx}{\columnwidth}{@{}ll >{\raggedright\arraybackslash}X @{}}
\toprule
\textbf{Currency} & \textbf{Encoding} & \textbf{Regex Pattern} \\
\midrule
BTC~\cite{nakamoto2008bitcoin, bip13, bip173}
  & Base58, Bech32
  & \texttt{bc1[0-9a-z]\{25,59\}|\allowbreak[13][a-km-zA-HJ-NP-Z1-9]\allowbreak\{25,34\}} \\
LTC~\cite{litecoin_addresses}
  & Base58, Bech32
  & \texttt{ltc1[0-9a-z]\{39,59\}|\allowbreak[LM3][a-km-zA-HJ-NP-Z1-9]\allowbreak\{26,33\}} \\
DASH~\cite{dash_address_conversion}
  & Base58
  & \texttt{[X7][1-9A-HJ-NP-Za-km-z]\allowbreak\{33\}} \\
ETH~\cite{wood2014ethereum, eip55}
  & Hexadecimal
  & \texttt{0x[0-9A-Fa-f]\{40\}} \\
XMR~\cite{vansaberhagen2013cryptonote, monero_address}
  & Base58
  & \texttt{4[0-9AB][1-9A-HJ-NP-Za-km-z]\allowbreak\{93\}|\allowbreak4[1-9A-HJ-NP-Za-km-z]\allowbreak\{105\}} \\
\bottomrule
\end{tabularx}
\end{table}

\subsection{On-Chain Validation}

Regex matching produces many false positives, including cryptographic hashes and other strings that resemble cryptocurrency addresses. To retain only valid addresses, we query the corresponding public blockchains and keep candidates that have participated in at least one confirmed transaction. We perform this validation using public blockchain APIs.\footnote{Blockchair: \url{https://api.blockchair.com/}; Etherscan: \url{https://api.etherscan.io/}. API queries performed on 11–12 May 2025.}

This process validates 27,131 cryptocurrency addresses. Bitcoin accounted for 23,882 (88.0\%) of these addresses, followed by Litecoin (3,084), Ethereum (155), and Dash (10). Monero addresses could not be validated because their transaction history is hidden by protocol-level privacy features.

\begin{figure}
    \centering
    \includegraphics[width=1\linewidth]{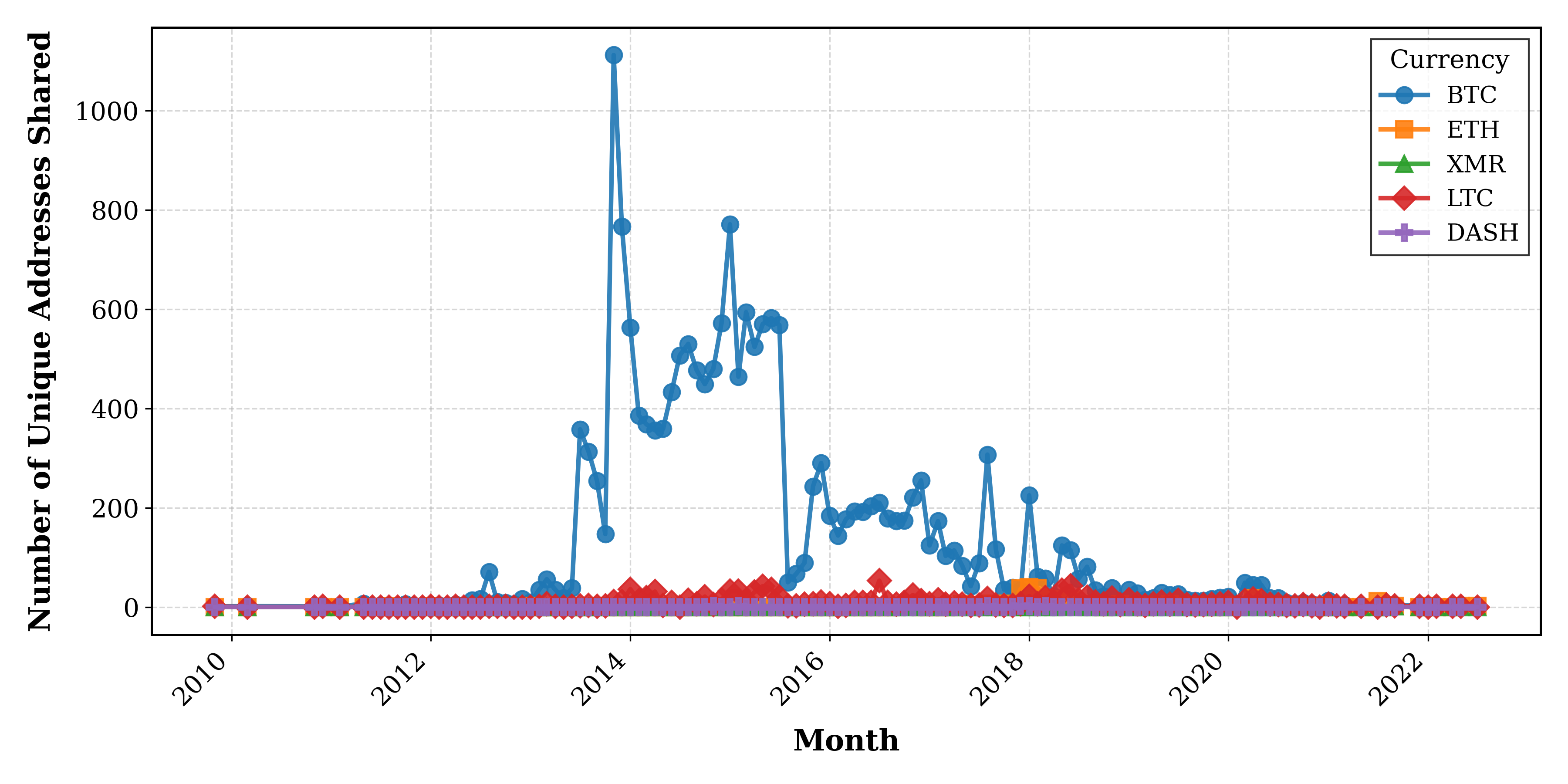}
  \caption{\textbf{Evolution of validated cryptocurrency addresses on HackForums (2010--2022).} Bitcoin historically dominates the dataset (88.0\%). Address sharing peaked around 2014, while the sharp decline post-2021 reflects the documented migration of cybercriminals from public forums to private messaging applications~\cite{paladini2024underground}, rather than a decrease in overall illicit activity.}
    \label{fig:evolution-btc}
\end{figure}

The predominance of Bitcoin is consistent with prior studies showing that it has historically been the primary cryptocurrency used in cybercrime~\cite{lee2019cybercriminal,foley2019sex}. Figure~\ref{fig:evolution-btc} illustrates the temporal distribution of the validated Bitcoin addresses. Given their overwhelming majority, we focus the remainder of the labeling pipeline on the 23,882 validated Bitcoin addresses.

\subsection{Two-Stage Labeling Pipeline}
\label{sec:labeling_pipeline}
Manually reviewing the context of all 23,882 validated Bitcoin addresses is impractical. We therefore adopt a two-stage labeling pipeline consisting of LLM-assisted candidate screening followed by manual expert verification. The LLM is used only to prioritize addresses for review; all final labels are assigned by a domain expert.

\paragraph{Stage 1: LLM-Assisted Screening.}
We used Qwen3-30B-A3B~\cite{yang2025qwen3}, a 30-billion-parameter Mixture-of-Experts language model (with 3 billion active parameters), served locally via Ollama on a single A100 node. For each address, we collected the forum posts referencing it, up to a maximum of~20 posts, prioritizing its earliest appearances in posts and contracts. To control prompt length, each post is truncated to a $\pm200$-character window around the address. The model is prompted in a zero-shot setting (temperature 0.3) to assign one of twelve cybercrime categories (Figure~\ref{fig:prompt}). Addresses assigned to any cybercrime category are retained as candidates for manual review, reducing the set from 23,882 to 2,606 addresses.

Table~\ref{tab:forum_categories} summarizes the distribution of the 2,438 expert-verified illicit Bitcoin addresses across the cybercrime taxonomy. Fraud (34.5\%) is the most prevalent category, followed by Malware (13.5\%), Investment Schemes (12.8\%), and Hacking Services (12.1\%). The LLM did not confidently classify any addresses as Counterfeit. Furthermore, addresses that the LLM labeled as 'Unclear' were discarded and not advanced to the expert verification stage.

\paragraph{Stage 2: Expert Verification.}
To ensure high label confidence, a domain expert manually reviewed all 2,606 candidate addresses flagged by the LLM. For each address, the expert examined every associated forum post together with the surrounding thread context and assigned one of three labels: \textit{illicit}, \textit{licit}, or \textit{excluded} when the available evidence was insufficient for a reliable decision. The expert verified only the binary illicit/licit label; the cybercrime category assigned during the LLM screening stage was retained. The review confirmed 2,438 illicit addresses, reclassified 98 as licit, and excluded 70 due to insufficient evidence, corresponding to an LLM screening precision of 93.6\%. 

\begin{table}[t]
\caption{Distribution of illicit addresses across categories, including labeling criteria and representative examples. \emph{Example Post Excerpt} entries are author-generated paraphrases that preserve the meaning and register of the underlying posts without reproducing them verbatim, in accordance with the CrimeBB Data Use Agreement (Section~\ref{sec:ethics}). No excerpt is a direct quotation or contains user identifiers, URLs, or contact handles.}

\label{tab:forum_categories}
\small
\centering
\begin{tabular}{@{}l r p{4.4cm} p{4.8cm}@{}}
\toprule
\textbf{Category} & \textbf{\# Addr. (Ratio)} & \textbf{Classification Criteria} & \textbf{Example Post Excerpt} \\
\midrule
Abuse           & 15 (0.62\%)   & Posts advertising illegal sexual or violent content.                          & \textit{[Excerpt withheld: advertises sexual content involving minors.]} \\
Account Selling & 82 (3.36\%)   & Compromised or stolen accounts for platforms such as Instagram, Netflix, or PayPal.           & \textit{``Selling verified PayPal and Coinbase accounts. BTC accepted.''} \\
Card Dumps      & 13 (0.53\%)   & Stolen card data, including CVVs, fullz, and track dumps.                      & \textit{``Fresh USA dump + PIN. Selling in bulk. Contact for price.''} \\
Drugs           & 11 (0.45\%)   & Narcotics, prescription medication, and synthetic drugs.                       & \textit{``Pure MDMA, next-day shipping, stealth packaging. BTC only.''} \\
Fraud & 841 (34.50\%) & Scams, fake documents, phishing kits, and identity theft.                     & \textit{``Full bank fraud tutorial. Includes spoof scripts and drops.''} \\
Investment      & 311 (12.76\%) & Ponzi schemes, fake crypto investment opportunities or pyramid schemes.                       & \textit{``Double your BTC in 24h. Limited slots available.''} \\
Laundering      & 161 (6.60\%)  & Money laundering, BTC mixers, or cash-out services.                           & \textit{``We clean dirty BTC. 15\% fee. Fast and anonymous.''} \\
Malware         & 329 (13.49\%) & Illicit tools including keyloggers, RATs, ransomware, and exploits.            & \textit{``New info stealer with FUD AV bypass. BTC only.''} \\
Membership      & 41 (1.68\%)   & Paid access to private marketplaces or forums.                                & \textit{``Selling invite to top-tier carding forum. Payment via BTC.''} \\
Services        & 296 (12.14\%) & Hacking-for-hire, custom phishing, DDoS attacks, and similar offerings.                & \textit{``Offering university grade changes. Remote access guaranteed.''} \\
Weapons         & 3 (0.12\%)    & Unlicensed firearms, explosives, or ammunition.                               & \textit{``Glock 19, no serial. Ships EU only. BTC required.''} \\
Other           & 335 (13.74\%) & Miscellaneous illicit offers not covered above.                               & \textit{``Selling private webcam feeds. BTC only.''} \\
\midrule
\textbf{Total}  & \textbf{2,438 (100.0\%)} & & \\
\bottomrule
\end{tabular}
\end{table}

\section{Address Characterization}
We collected 23,882 validated Bitcoin addresses shared on HackForums between 2009 and 2024, including 2,438 expert-verified illicit addresses. Table~\ref{tab:onchain_stats} presents aggregate statistics for the validated Bitcoin addresses. Compared with the remaining validated addresses, illicit addresses exhibit higher on-chain activity, with median received and sent volumes approximately ten times larger (0.1737 BTC vs. 0.0180 BTC received), suggesting that addresses associated with illicit services typically process substantially larger transaction flows.

\subsection{Longitudinal Evolution}

\begin{figure}
    \centering
    \includegraphics[width=1\linewidth]{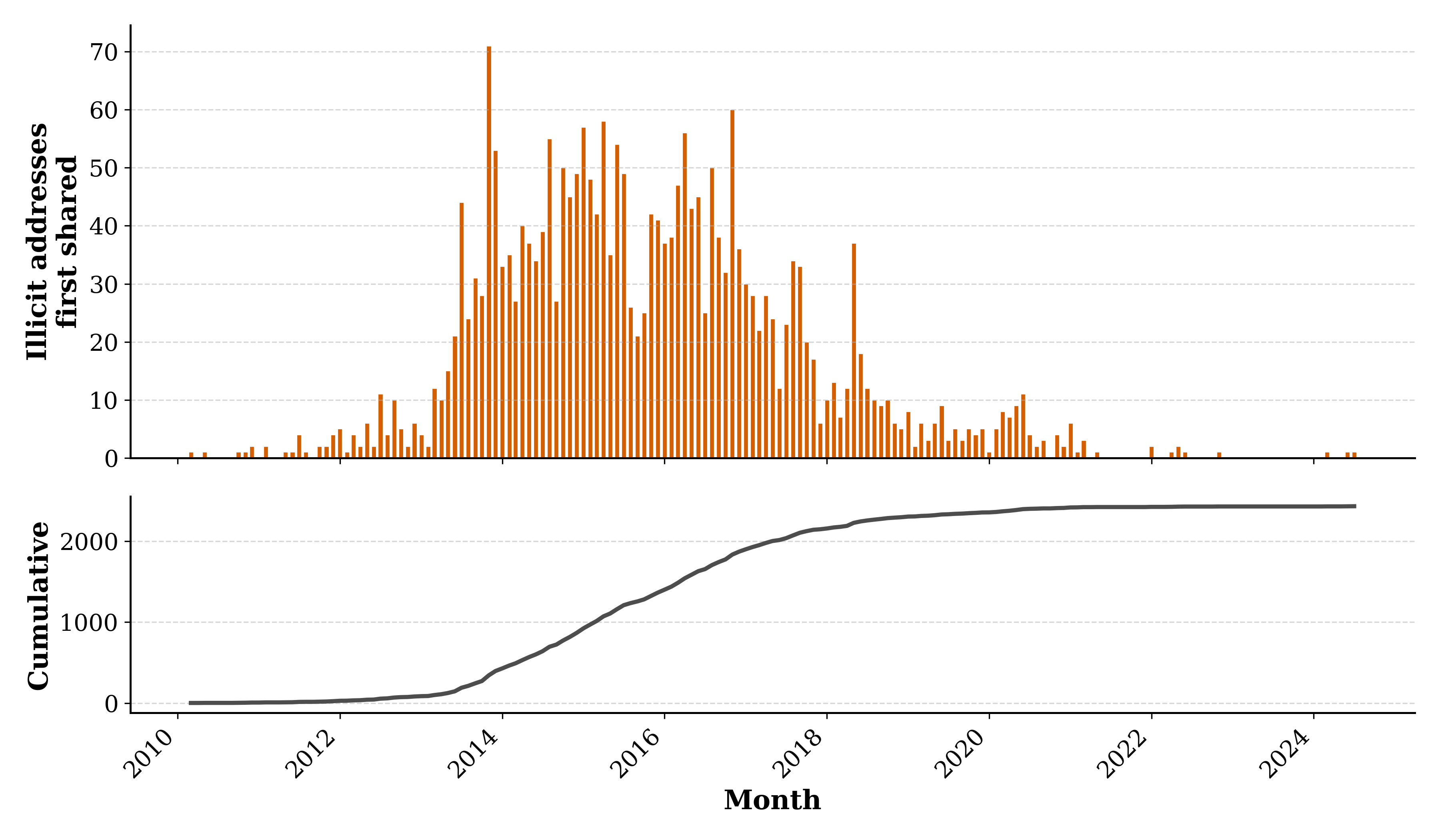}
    \caption{Temporal evolution of verified illicit Bitcoin addresses on HackForums (2010–2024)}
    \label{fig:evolution-illicit}
\end{figure}

Figures~\ref{fig:evolution-btc} and~\ref{fig:evolution-illicit} illustrate the temporal evolution of Bitcoin address sharing on HackForums. Although the forum snapshot extends to September~2024, the number of newly shared, valid cryptocurrency addresses drops to nearly zero after 2022. Consequently, Figure~\ref{fig:evolution-btc} primarily reflects activity through 2022. In contrast, Figure~\ref{fig:evolution-illicit} extends to 2024 because it is based on the last on-chain transaction date of each verified illicit address rather than the forum posting date. Figure~\ref{fig:evolution-btc} presents the annual distribution of all 23,882 validated Bitcoin addresses, whereas Figure~\ref{fig:evolution-illicit} focuses on the 2,438 expert-verified illicit addresses. Its upper panel shows the monthly number of newly observed illicit addresses, and the lower panel reports the cumulative growth of the dataset over time.

Both figures reveal a rapid increase in address sharing beginning in late 2013, followed by a pronounced peak in 2014. This period coincides with the emergence of early crypto-ransomware such as CryptoLocker (active from September~2013 to early~2014), which established Bitcoin as a practical payment mechanism for ransomware extortion~\cite{liao2016behind}, as well as the broader growth of Bitcoin adoption within the cybercrime ecosystem~\cite{foley2019sex}. Elevated activity persists through 2017, after which the number of newly shared addresses declines steadily.

This decline is consistent with well-documented changes in underground operational practices~\cite{paladini2024underground}, where public forums increasingly serve as advertising channels while payment negotiations shift to encrypted platforms such as Telegram~\cite{3766078.3766327}. HackForums captures a progressively smaller fraction of illicit cryptocurrency activity in recent years, despite the continued evolution and expansion of the broader cybercrime ecosystem.

\begin{table}
\centering
\caption{On-chain and forum statistics for validated Bitcoin addresses.}
\label{tab:onchain_stats}

\begin{tabular}{lrrr}
\toprule
\textbf{Metric} & \textbf{All Validated} & \textbf{Illicit} & \textbf{Unknown} \\
\midrule
Total Addresses & 23{,}882 & 2{,}438 & 21{,}444 \\
First forum appearance & 2009-10-10 & 2010-03-30 & 2009-10-10 \\
Last forum appearance & 2024-07-07 & 2024-07-07 & 2024-05-13 \\
Total mentions & 452{,}950 & 50{,}507 & 402{,}443 \\
Tx count (median) & 4 & 7 & 4 \\
BTC received (median) & 0.0217 & 0.1737 & 0.0180 \\
BTC sent (median) & 0.0209 & 0.1616 & 0.0172 \\
First on-chain Tx & 2009-01-03 & 2011-03-05 & 2009-01-03 \\
\bottomrule
\end{tabular}%

\end{table}

\begin{figure}[t]
\centering
\begin{tcolorbox}[
    colback=gray!5!white,
    colframe=gray!75!black,
    title=\textbf{Prompt -- Illicit Address Classification},
    fontupper=\small\sffamily,
    boxrule=0.8pt,
    left=4pt, right=4pt, top=4pt, bottom=4pt
]
\textbf{[Role]}\\
You are an expert digital forensics investigator specializing in
cryptocurrency-facilitated cybercrime. Your task is to analyze unstructured text
from underground forum posts to identify illicit goods or services associated
with a specific cryptocurrency address.\\

\textbf{[Input Data]}
\begin{itemize}[leftmargin=*, nosep]
    \item \textbf{Thread Title:} \{thread\_name\}
    \item \textbf{Post Content:} \{content\}
    \item \textbf{Target Address:} \texttt{\{address\}}
\end{itemize}

\textbf{[Taxonomy]}\\
Classify the activity into exactly one of the following categories:
\begin{itemize}[leftmargin=*, nosep]
    \item \textbf{Abuse:} Sexual content, violent imagery, or human trafficking.
    \item \textbf{Account Selling:} Compromised credentials, social media accounts, or invite codes.
    \item \textbf{Counterfeit:} Forged documents (IDs, passports) or counterfeit currency.
    \item \textbf{Card Dumps:} Stolen credit card details (CVV, fullz) or banking logs.
    \item \textbf{Drugs:} Narcotics, controlled substances, or pharmaceuticals.
    \item \textbf{Fraud:} Scam-as-a-service, social engineering, or identity theft.
    \item \textbf{Investment:} Ponzi schemes, High-Yield Investment Programs (HYIPs), or fraudulent investment platforms.
    \item \textbf{Laundering:} Mixing services, cash-out schemes, or money laundering.
    \item \textbf{Malware:} Ransomware, RATs, exploits, or botnet rentals.
    \item \textbf{Membership:} Access to private forums, VIP groups, or databases.
    \item \textbf{Services:} Illicit tasks (e.g., physical harm, hacking-for-hire, DDoS).
    \item \textbf{Weapons:} Firearms, explosives, or ammunition.
    \item \textbf{Other:} Illicit goods not covered by the above categories.
    \item \textbf{Unclear:} Insufficient context to determine the nature of the address.
\end{itemize}

\textbf{[Constraints]}
\begin{enumerate}[leftmargin=*, nosep]
    \item Analyze the context of the \texttt{\{address\}} within the post.
    \item Ignore secondary discussions; focus on the primary item or service for sale.
    \item Output format: return a JSON object \texttt{\{"category": <label>\}}, where \texttt{<label>} is exactly one taxonomy name, with no explanation.
\end{enumerate}
\end{tcolorbox}
\caption{The structured prompt template used to categorize illicit Bitcoin
addresses based on forum context.}
\label{fig:prompt}
\end{figure}

\subsection{Evidence Corroboration}
Forum discussions sometimes contain multiple references to the same Bitcoin address, providing additional contextual evidence about its use. Table~\ref{tab:corroboration} groups the expert-verified illicit addresses into three corroboration profiles. Reported addresses are referenced by multiple independent users, typically in scam or abuse reports. Self-advertised addresses are repeatedly posted by the same user across multiple threads to promote illicit goods or services. Single-mention addresses appear in only one thread and therefore lack corroboration from additional forum occurrences.

Across the seven named categories (2,061 addresses), 837 addresses (40.6\%) appear in multiple forum contexts: 512 are self-advertised across multiple threads by the same user, while 325 are reported by multiple independent users. Fraud has the highest proportion of victim-reported addresses (185 addresses), reflecting users returning to warn others after being scammed. In contrast, Investment schemes and Malware exhibit the highest rates of self-advertised addresses, as operators repeatedly post the same payment address to promote their services across multiple threads.

\begin{table}[t]
\centering
\caption{Corroboration profiles of HackForums illicit addresses by category. \emph{Self-advertised}: $\geq$2 threads, 1 reporter. \emph{Reported}: $\geq$2 independent reporters. \emph{Single-mention}: 1 thread, 1 reporter.}
\label{tab:corroboration}
\small
\begin{tabular}{lrrrr}
\toprule
Category & Total & Self-adv. & Reported & Single \\
\midrule
Fraud          & 841 &  86 (10.2\%) & 185 (22.0\%) & 570 (67.8\%) \\
Investment     & 311 & 156 (50.2\%) &  29  (9.3\%) & 126 (40.5\%) \\
Malware        & 329 & 134 (40.7\%) &  38 (11.6\%) & 157 (47.7\%) \\
Service        & 296 &  91 (30.7\%) &  29  (9.8\%) & 176 (59.5\%) \\
Laundering     & 161 &  32 (19.9\%) &  19 (11.8\%) & 110 (68.3\%) \\
Acct.\ selling &  82 &  11 (13.4\%) &  13 (15.9\%) &  58 (70.7\%) \\
Membership     &  41 &   2  (4.9\%) &  12 (29.3\%) &  27 (65.9\%) \\
\bottomrule
\end{tabular}
\end{table}

\subsection{Estimating Screening Recall}
\label{sec:recall}

The expert verification stage reviewed only the 2,606 Bitcoin addresses flagged by the LLM as potentially illicit. To estimate the recall of this screening step, we manually evaluated a sample of the addresses that the LLM did not flag. From the 21,276 validated Bitcoin addresses classified as non-illicit by the LLM, we selected a random sample of 200. The domain expert reviewed the forum context for each address.

Among the 200 sampled addresses, 8 were identified as illicit, corresponding to an estimated false-negative rate of 4.0\%. Extrapolating this point estimate to the 21,276 unflagged addresses suggests approximately 851 missed illicit addresses and an estimated screening recall of 74.1\%. Because this estimate is based on a sample of 200 addresses, it should be interpreted as an approximate measure of screening coverage rather than an exact recall estimate.

The 8 missed addresses fall into two confidence tiers. Four are classified as \textit{firm}: the post text contains explicit transactional evidence linking the address to a named cybercrime product or service. For example, \texttt{1Giqo\dots pm5} is advertised as a payment destination for a malware-spreading method sold via an autobuy. Address \texttt{39h5C\dots Hp2} appears in a payment receipt for a one-month MailMurder subscription. The other two firm cases involve a payment for the TOPHER crypter (\texttt{3BBoM\dots otu}) and a solicitation to fund the development of an Orcus RAT variant (\texttt{1QJRf\dots dn4}).

The remaining four addresses are classified as \textit{arguable}: the link to illicit activity relies on contextual inference rather than a direct transactional statement. For instance, \texttt{1Da92\dots Wr} confirms a purchase in a thread advertising spreading methods, but the post does not describe the payload. Address \texttt{1Mi9t\dots RS} belongs to a pentesting service that is cross-promoted with a doxing service through a persistent forum signature. The distinction between tiers is whether the illicit nature of the transaction is stated explicitly in the post or must be inferred from surrounding context, such as thread titles, co-advertisements, or dispute narratives.

These missed cases indicate that the LLM screening stage primarily
under-detects addresses whose illicit context is implicit or embedded in
forum-specific conventions, such as signature blocks and dispute threads, rather
than explicitly described in the post. Despite this limitation, the estimated
recall of roughly 74\% indicates that zero-shot screening
captured the majority of illicit addresses while reducing expert review from
hundreds of thousands of forum posts to a high-precision candidate set with a
verified precision of 93.6\%.

\section{Discussion and Limitations}

Our dataset provides evidence-backed Bitcoin labels, but the methodology has several limitations.

\textbf{Extraction Heuristics.} The initial address extraction relies on broad regular expressions rather than protocol-specific cryptographic checksums (e.g., Base58Check for legacy Bitcoin addresses or Bech32 for SegWit). Some regex patterns also overlap: for instance, legacy P2SH Bitcoin addresses and legacy Litecoin addresses both begin with the prefix \texttt{3}. We mitigated these ambiguities downstream by requiring a successful resolution against the respective blockchain's public API. This step filtered out malformed strings and cross-chain collisions.

\textbf{LLM Screening Bias.} Our manual sampling in Section~\ref{sec:recall} estimates a recall of 74.1\% for the LLM screening stage. However, the model relies on zero-shot, English-dominant training data. As a result, our dataset likely under-samples illicit activity negotiated in heavily obfuscated slang or non-English languages. Fine-tuned models trained on underground forum dialects could improve recall in future work.

\textbf{Annotation Constraints.} The manual verification was conducted by a single domain expert. Because we did not employ multiple annotators, we cannot report an inter-rater reliability score (e.g., Cohen's Kappa). In addition, the expert verified only the binary illicit or licit label for each address. The specific sub-category (e.g., \textit{Fraud} vs.\ \textit{Malware}) was retained directly from the LLM's zero-shot classification; these categories may therefore contain occasional misassignments or inconsistencies introduced by the model.

\section{Ethics and Data Handling}
\label{sec:ethics}
We conducted this study on publicly posted content and took several steps to
limit harm.

\paragraph{Data handling and DUA.} We obtained the forum dataset via the CrimeBB academic archive~\cite{Pastrana2018CrimeBB}, provided by the Cambridge Cybercrime Centre. We adhere strictly to their Data Use Agreement (DUA), which prohibits the redistribution of raw, deanonymized post text or user profiles. All data processing occurred in secure, local environments. 

\paragraph{Human subjects.}  We designed this study to be entirely non-interventional. Our analysis is based strictly on archived, publicly visible forum posts; we did not interact with forum users, create accounts for active scraping, or engage in any transactions.

\paragraph{Sensitive content.} During manual annotation, the expert reviewer
encountered posts advertising illegal goods and services, including a small
number of posts referencing sexual exploitation. To minimize psychological harm
and avoid any possibility of acquiring illicit material, verification was
strictly limited to reading archived, plain-text post content. The expert did
not follow hyperlinks, execute attachments, resolve darkweb domains, or attempt
to retrieve any media. No image, video, or file content was accessed at any
point in this study.

\paragraph{Responsible release.} Bitcoin addresses and their on-chain
transactions are already public. The released artifact contains only addresses,
their category labels, and temporal metadata; it contains no user identifiers
and no post text. The short excerpts shown in Table~\ref{tab:forum_categories}
appear only in this paper, for illustration, and are not part of the released
dataset.

\section{Conclusion}
We presented a pipeline for extracting, validating, and labeling illicit Bitcoin addresses from underground forums and applied it to more than 42 million HackForums posts. The resulting dataset contains 2,438 expert-verified illicit Bitcoin addresses spanning 12 cybercrime categories. Our pipeline combines LLM-based candidate screening with expert verification, enabling efficient large-scale annotation while grounding every label in contextual forum evidence and on-chain activity. Unlike existing datasets that often rely on community reports or proprietary intelligence, our dataset provides transparent label provenance and a fully reproducible construction process. Our analysis confirms Bitcoin's historical role as the dominant payment mechanism for cybercrime and reveals a marked decline in public forum address sharing after 2021, consistent with a shift toward private communication channels. We publicly release the dataset and the complete extraction pipeline to support future research in blockchain analytics, illicit address detection, and cyber threat intelligence.

\bibliographystyle{plainurl}
\bibliography{sample}

\end{document}